\documentclass[structabstract]{aa}  
\usepackage{graphicx}
\usepackage{array}
\usepackage{txfonts}
\usepackage{natbib}
\bibpunct{(}{)}{;}{a}{}{,}

\begin{document}
   \title{Multiwavelength campaign on Mrk 509}
   \subtitle{XIV. Chandra HETGS spectra}

\author{J.S. Kaastra\inst{1,2,3}
  \and J. Ebrero\inst{4,1}
  \and N. Arav\inst{5}
  \and E. Behar\inst{6}
  \and S. Bianchi\inst{7}
  \and G. Branduardi-Raymont\inst{8}
  \and M. Cappi\inst{9}
  \and E. Costantini\inst{1}
  \and G.A. Kriss\inst{10,11}
  \and B. De Marco\inst{12}
  \and M. Mehdipour\inst{8,1}
  \and S. Paltani\inst{13}
  \and P.-O. Petrucci\inst{14,15}
  \and C. Pinto\inst{16}
  \and G. Ponti\inst{12}
  \and K.C. Steenbrugge\inst{17,18}
  \and C.P. de Vries\inst{1}
  }
  
\institute{SRON Netherlands Institute for Space Research, Sorbonnelaan 2,
           3584 CA Utrecht, the Netherlands 
	   \and
	   Department of Physics and Astronomy, Universiteit Utrecht, 
	   P.O. Box 80000, 3508 TA Utrecht, the Netherlands
	   \and
	   Leiden Observatory, Leiden University, PO Box 9513, 
	   2300 RA Leiden, the Netherlands
	   \and
	   European Space Astronomy Centre (ESAC), P.O. Box 78, E-28691
	   Villanueva de la Ca\~nada, Madrid, Spain.
	   \and
	   Department of Physics, Virginia Tech, Blacksburg, VA 24061, USA
	   \and
	   Department of Physics, Technion-Israel Institute of Technology, 
	   Haifa 32000, Israel 
	   \and 
	   Dipartimento di Matematica e Fisica, Universit\`a degli Studi Roma Tre, 
	   Via della Vasca Navale 84, 00146 Roma, Italy 
	   \and
	   Mullard Space Science Laboratory, University College London, 
	   Holmbury St. Mary, Dorking, Surrey, RH5 6NT, UK
           \and
	   INAF-IASF Bologna, Via Gobetti 101, 40129 Bologna, Italy
	   \and
	   Space Telescope Science Institute, 3700 San Martin Drive, 
	   Baltimore, MD 21218, USA
	   \and
	   Department of Physics and Astronomy, The Johns Hopkins University,
	   Baltimore, MD 21218, USA
           \and
	   Max Planck Institut f\"ur Extraterrestrische Physik, 
	   D-85741 Garching, Germany
	   \and
	   University of Geneva, 16, ch. d'Ecogia, 1290 Versoix, Switzerland 	
	   \and
	   Univ. Grenoble Alpes, IPAG, F-38000 Grenoble, France
	   \and
	   CNRS, IPAG, F-38000 Grenoble, France
	   \and
           Institute of Astronomy, University of Cambridge, Madingley Road, Cambridge CB3 0HA, UK
	   \and
           Instituto de Astronom\'ia, Universidad Cat\'olica del Norte, 
	   Avenida Angamos 0610, Casilla 1280, Antofagasta, Chile
	   \and
	   Department of Physics, University of Oxford, Keble Road, 
	   Oxford OX1 3RH, UK
} 
\date{\today}

\abstract
{We present in this paper the results of a 270~ks Chandra HETGS observation in
the context of a large multiwavelength campaign on the Seyfert galaxy Mrk~509.}
{The HETGS spectrum allows us to study the high ionisation warm absorber and
the Fe-K complex in Mrk~509. We search for variability in the spectral
properties of the source with respect to previous observations in this campaign,
as well as for evidence of ultra-fast outflow signatures.}
{The Chandra HETGS X-ray spectrum of Mrk~509 was analysed using the
\textsl{SPEX} fitting package.}
{We confirm the basic structure of the warm absorber found in the 600~ks
XMM-Newton RGS observation observed three years earlier, consisting of five
distinct ionisation components in a multikinematic regime. We find little or no
variability in the physical properties of the different warm absorber phases
with respect to previous observations in this campaign, except for component D2
which has a higher column density at the expense of component C2 at the same
outflow velocity ($-240$~km\,s$^{-1}$). Contrary to prior reports we find no
$-700$~km\,s$^{-1}$ outflow component. The \ion{O}{viii} absorption line
profiles show an average covering factor of $0.81\pm 0.08$ for outflow
velocities faster than $-100$~km\,s$^{-1}$, similar to those measured in the UV.
This supports the idea of a patchy wind. The relative metal abundances in the
outflow are close to proto-solar. The narrow component of the Fe K$\alpha$
emission line shows no changes with respect to previous observations which
confirms its origin in distant matter. The narrow line has a red wing that can
be interpreted to be a weak relativistic emission line. We find no significant
evidence of ultra-fast outflows in our new spectrum down to the  sensitivity
limit of our data.}
{}

\maketitle

\section{Introduction}

Active galactic nuclei (AGN) are powered by gravitational accretion of matter
onto a central supermassive black hole. It is thought that the intense radiation
field might be powerful enough to drive gas outflows from the nucleus. The
widespread occurrence of these outflows was recognised after the development of
high-resolution UV and X-ray spectrographs, which were able to identify their
signatures in the form of absorption lines of a photoionised gas, the so-called
warm absorber (WA), blueshifted with respect to the rest frame of the source
\citep{crenshaw1999,kaastra2000}. An extensive overview of these outflows can be
found in \citet{crenshaw2003}.

Despite their importance, several questions regarding the origin, geometry, and
structure of the WA remain open. The proposed origins for AGN outflows are
disc-driven winds \citep{elvis2000} and thermally driven winds emanating from
the putative dusty torus \citep{krolik2001}. The discovery of blueshifted Fe
K-shell lines in radio-quiet AGN \citep[e.g.][]{chartas2002,chartas2003} were
interpreted with the so-called ultra-fast outflows, defined as highly ionised
winds ($\log \xi \sim 3 - 6$) outflowing at velocities faster than
$-10\,000$~km\,s$^{-1}$ \citep{tombesi2010}. In this context, it has been
proposed that the ultra-fast outflows are launched as a disc wind, close to the
central black hole, whereas WA are launched farther out, thus forming a
stratified wind \citep{kazanas2012,tombesi2013}.

Here we present the results of a 270~ks Chandra HETGS observation of the Seyfert
1 galaxy Mrk 509 in the context of a large multiwavelength campaign. Mrk 509 is
one of the brightest Seyfert galaxies in the sky ($L\,(1-1000~{\rm Ryd})=3.2
\times 10^{38}$~W), and it is considered one of the closest ($z = 0.034397$)
Seyfert 1/QSO hybrids. Because of its brightness and confirmed presence of UV
and X-ray absorbers, Mrk 509 was the subject of an extensive multiwavelength
campaign in 2009 described in \citet{kaastra2011a}. The present Chandra data
were taken in 2012.

This paper is organised as follows. In Sect.~\ref{sect:reduction} we describe
the data reduction. In Sect.~\ref{sect:lineanalysis} we analyse the spectral
lines, while in Sect.~\ref{sect:o8prof} we study the \ion{O}{viii} line profile.
The overall modelling of the outflow and the Fe-K complex are described in
Sect.~\ref{sect:outflow} and Sect.~\ref{sect:fek}, respectively. In
Sect.~\ref{sect:ufos} we search for ultra-fast outflows in our spectrum. In
Sect.~\ref{sect:discussion} we discuss our results and, finally, we report our
conclusions in Sect.~\ref{sect:conclusions}.

Throughout the paper, the C-statistics fitting method is used, and the quoted
errors refer to 68.3\% confidence level ($\Delta C = 1$ for one parameter of
interest) unless otherwise stated.

\section{Data reduction}\label{sect:reduction}

Mrk~509 was observed by Chandra with the HETGS between 4 and 9 September 2012.
The Chandra observation was split into two parts, separated by 19 hours:
observation nrs. 13864 and 13865, with exposure times of 170 and 99~ks,
respectively. The data were analysed using the standard CIAO version 4.4 tools.

The spectra and response matrices were converted into SPEX \citep{kaastra1996}
format, and that package was used for all spectral analysis. The standard High
Energy Grating (HEG) and Medium Energy Grating (MEG) spectra are
slightly-oversampled hence we binned them by a factor of 2 to about 1/3--1/2 of
the full width at half maximum (FWHM) of the respective gratings. 

\subsection{Spectral modelling}

It appears that when the fluxed MEG and HEG spectra are plotted, the HEG/MEG
flux ratio is fairly constant at an average value of $0.954\pm 0.004$ in the
3--13~\AA\ band. In the lower wavelength band, between 1.5--3~\AA, this ratio is
rising with an average value of $0.999\pm 0.016$. At the other side of the
spectrum, between 13--16~\AA, the HEG/MEG flux ratio is lower, on average
$0.90\pm 0.05$. For these reasons we restricted the data range to 2.5--26~\AA\
for MEG (at the long wavelength end the statistics becomes too poor), and to
1.55--15.5~\AA\ for HEG. In our combined spectral fits we rescaled the response
matrix of the HEG uniformly by a factor of 0.954 to get consistent spectra for
HEG and MEG over the relevant wavelength range.

\subsection{Time variability}

We checked variability of the source by constructing the zeroth order lightcurve
using 1000~s bins. For the first observation, the average count rate was
constant at a value of $0.3351\pm 0.0014$~counts/s. We find strict upper limits
to any linear flux changes; they correspond to flux differences compared to the
mean flux of less than 0.7\%. Interestingly, the second observation has the same
average count rate within the error bars, $0.3343\pm 0.0018$~counts/s. However,
this observation shows a weak declining trend, with an almost linear flux
decrease of $4.6\pm 1.9$\% between the start and the end of the observation.
Compared to the XMM-Newton observations 3 years earlier, the variability on
these short timescales is remarkably low. For this reason we have combined both
observations and only consider the single time-averaged HETGS spectrum.

Apart from the flux constancy during the Chandra observations in 2012, also the
fluxed continuum spectrum of the Chandra observation over the 5--25~\AA\ range
agrees within 5\% with the fluxed RGS spectrum taken in 2009. The similarity is
even stronger: we have measured the equivalent width of the seven strongest
absorption lines in common between the RGS and HETGS spectra, and found that
they are the same within the error bars. The HST/COS UV band also shows no
strong differences in flux between the 2009 and 2012 observations (less than 5\%
at 1360~\AA). Given this similarity of the spectra, we will adopt for our
modelling the same ionising spectral energy distribution (SED) as used for the
analysis of the XMM-Newton data of 2009 \citep{kaastra2011a,detmers2011}.

\subsection{Velocity scale}

The proper wavelength calibration for the RGS spectra of Mrk~509 was rather
complex, see \citet{kaastra2011b}. This is essentially due to the lack of a
zeroth-order spectrum in the RGS. Because of small non-linearities in the HRC-S
detector used with the Chandra LETGS spectra in 2009 there remains some
uncertainty in the wavelength scale, estimated by \citet{kaastra2011b} to be
about 1.8~m\AA, corresponding to about 30~km\,s$^{-1}$.

For the HETGS the situation is more straightforward because of the presence of
the zeroth-order combined with the extremely linear wavelength scale provided by
the solid state ACIS-S detector chips. We have corrected our standard extracted
HETGS spectra only for the Doppler motion of the Earth around the Sun, which was
12.7 km\,s$^{-1}$ towards Mrk~509. 

We have measured the outflow velocities derived from the 7 strongest absorption
lines in common between the HETGS and the RGS spectra (\ion{Ne}{ix} and
\ion{Ne}{x} 1s--2p; \ion{O}{vii} 1s--2p, 1s--3p and 1s--4p, and \ion{O}{viii}
1s--2p and 1s--3p). There is a perfect linear correlation between both velocity
scales. Adopting the HETGS velocity scale to be correct, the velocities derived
from the RGS spectra in \citet{kaastra2011b} and subsequent papers need to be
increased by $+55\pm 21$~km\,s$^{-1}$ (i.e. they are less blueshifted). This
correction is just about twice the uncertainty of 30~km\,s$^{-1}$ that was
derived before on this for the RGS velocity scale. 

From the above analysis we have excluded one line, namely the \ion{Mg}{xi}
1s--2p line. The RGS spectrum \citep{kaastra2011b} showed a clear line
($>5\sigma$ significance) at a velocity of $-545\pm 150$~km\,s$^{-1}$, corrected
to the HETGS scale. However, for the same \ion{Mg}{xi} transition the HETGS
spectrum has a velocity of $-200\pm 60$~km\,s$^{-1}$. We discuss this in
Section~\ref{sect:discussion}. From now onwards we will report all velocities
from the RGS analysis, including those reported by \citet{detmers2011},
corrected to the HETGS scale. 

\section{Analysis of spectral lines}\label{sect:lineanalysis}

\begin{figure}[!tbp]
\resizebox{\hsize}{!}{\includegraphics[angle=0]{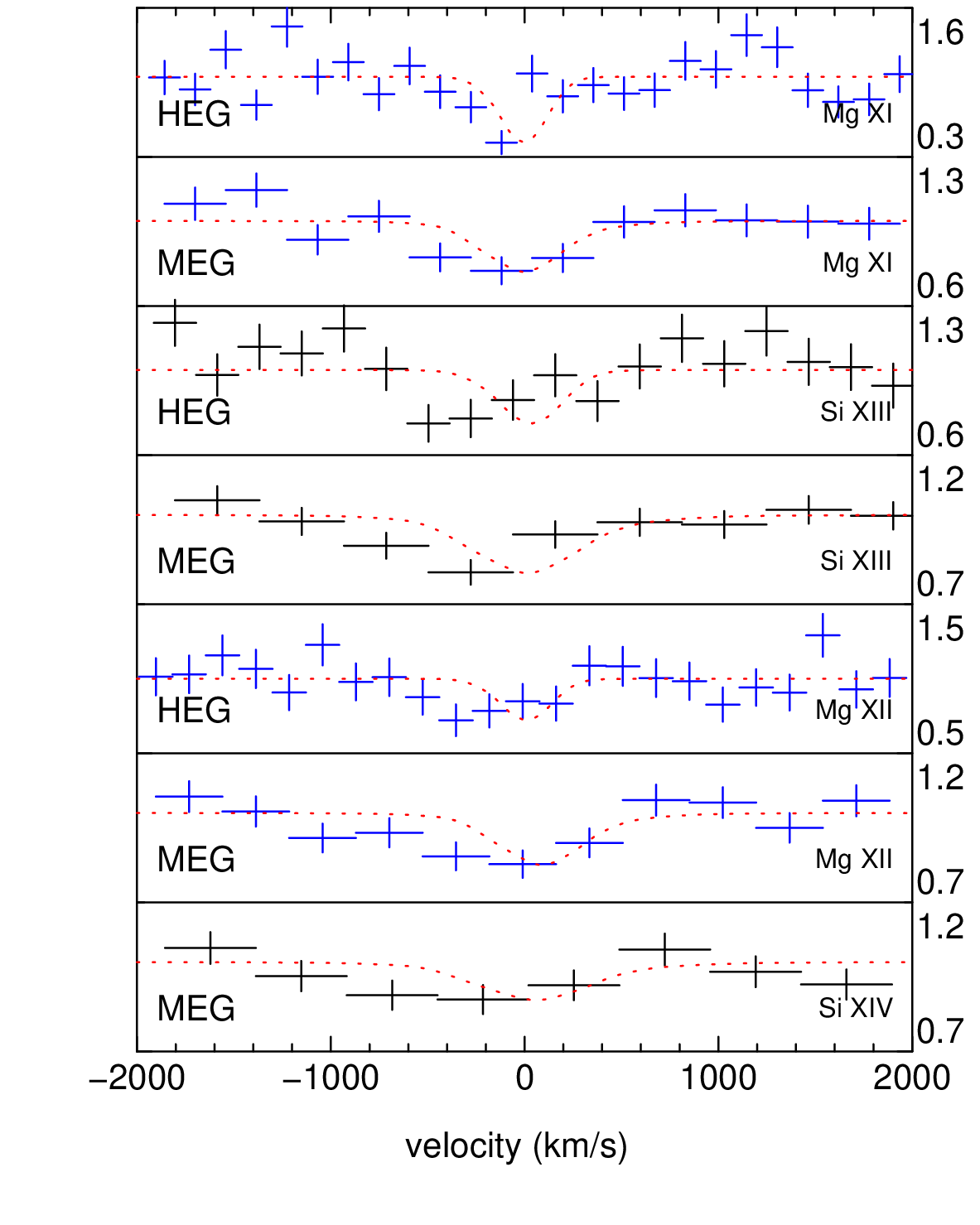}} 
\caption{Normalised line profiles for Si and Mg lines. Line identifications and
gratings are indicated in the panels. The dotted lines correspond to the
expected line profile for an infinitely narrow absorption line at outflow
velocity 0 convolved with the instrumental line-spread function and with a depth
corresponding to the depth of the observed spectral line. They serve to show the
typical instrumental broadening.}
\label{fig:simg}
\end{figure}

\begin{figure}[!tbp]
\resizebox{\hsize}{!}{\includegraphics[angle=0]{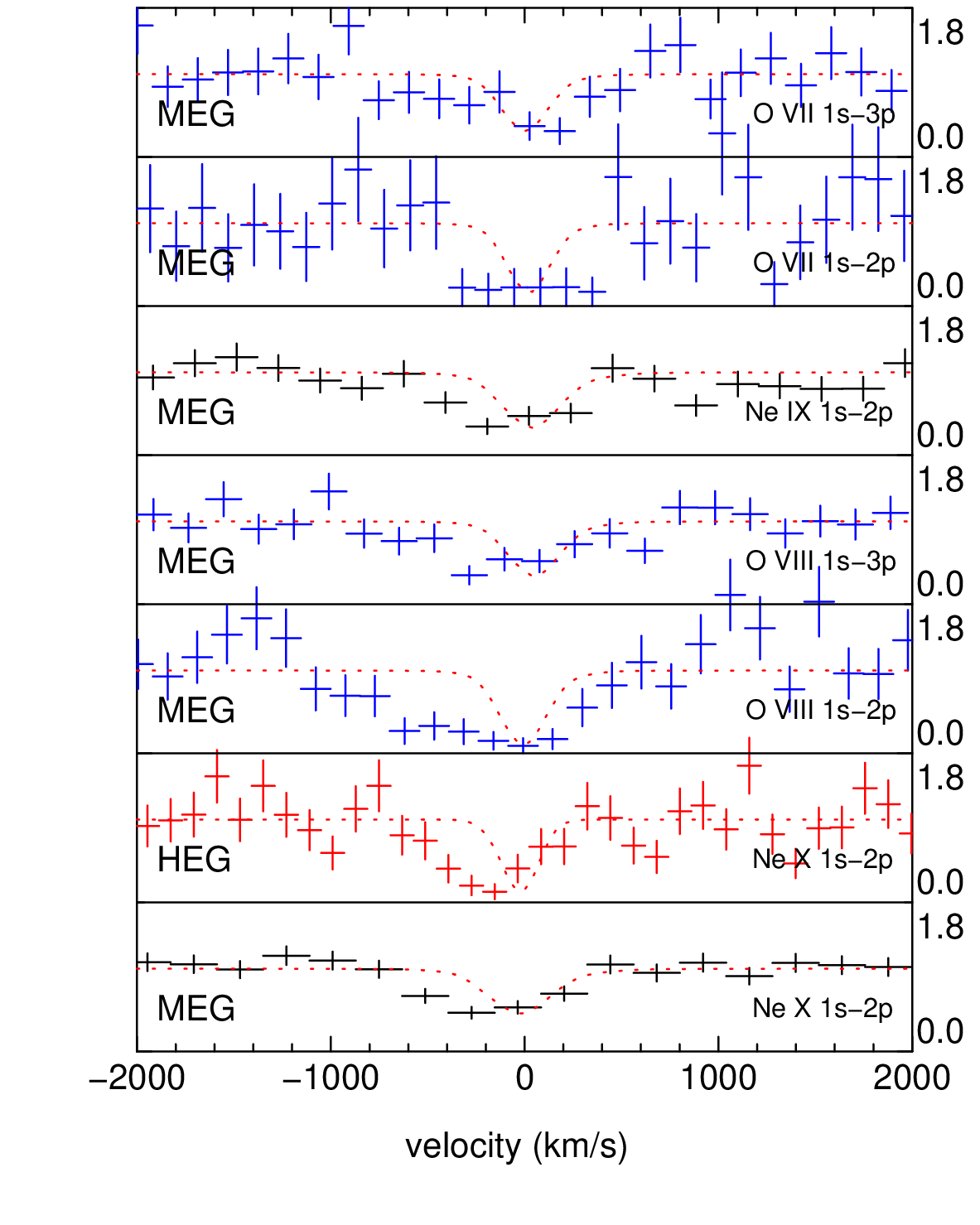}}
\caption{As Fig.~\ref{fig:simg}, but for Ne and O lines. }
\label{fig:neo}
\end{figure}

We have determined line profiles for the stronger spectral lines as follows.
Over small wavelength ranges, the spectrum expressed in
photons\,m$^{-2}$\,s$^{-1}$\,\AA$^{-1}$ is rather flat. Therefore, we determined
the average continuum flux level for each line in a band between 10\,000 and
20\,000 km\,s$^{-1}$ offset at each side of the line. From this continuum
range we exclude the narrow absorption lines as well as the weak broad emission
lines associated with some of these lines \citep{detmers2011}. The spectra near
the lines were normalised to this level and are shown in Figs.~\ref{fig:simg}
and \ref{fig:neo}.

It is seen that all lines are significantly broader than the spectral resolution
of the instrument, and most of them show a blue wing extending up to several
hundred km\,s$^{-1}$, a clear signature of outflow. Several spectral lines show
almost black line cores, indicative for a high optical depth, in particular the
\ion{Ne}{x} and \ion{O}{viii} 1s--2p transitions. This last line even shows some
evidence of a red wing, and in its blue wing extends up to $-600$~km\,s$^{-1}$
or even up to $-1200~$km\,s$^{-1}$ (Fig.~\ref{fig:neo}). Interestingly, while
the line is very deep, it appears not to be completely black. A similar
behaviour appears to be the case for the \ion{O}{vii} 1s--2p line, although
because of the strong drop of sensitivity of the MEG towards long wavelengths
the significance is somewhat lower than for \ion{O}{viii}.

Because for \ion{O}{viii} both the 1s--2p and 1s--3p lines are well-resolved and
have a good statistical quality, we first perform a detailed analysis of the
line profiles of both lines.

\section{Velocity-resolved line spectroscopy of
\ion{O}{viii}}\label{sect:o8prof}

In the UV-band, thanks to the high-resolution of instruments like FUSE, STIS and
COS, it has been possible to do velocity-resolved line spectroscopy on sets of
lines from the same ion in order to derive covering factors and column densities
as a function of outflow velocity \citep[e.g.][]{arav2002}. In the X-ray band
this has been harder to do because of the lower spectral resolution and smaller
number of photons. Mrk~509 is one of the brightest Seyferts on the sky, however,
and our deep exposure with the HETGS provides us with a suitable pair of lines
from \ion{O}{viii}: the 1s--2p (Ly$\alpha$) and 1s--3p (Ly$\beta$) lines
(Fig.~\ref{fig:neo}). The 1s--2p line is several times broader than the spectral
resolution of the MEG grating. 

We have modelled the velocity-dependent transmission $T(\varv)$ of both
components as follows:

\begin{equation}
T_\alpha(\varv) = 1 - f_{\rm c}(\varv) + f_{\rm c}(\varv) {\rm
e}^{\displaystyle{-\tau_\alpha(\varv)}},
\end{equation}
\begin{equation}
T_\beta(\varv) = 1 - f_{\rm c}(\varv) + f_{\rm c}(\varv) {\rm
e}^{\displaystyle{-\tau_\beta(\varv)}}.
\end{equation}
Here $\tau_\beta(\varv) = (\lambda_\beta f_\beta/\lambda_\alpha f_\alpha)\,
\tau_\alpha(\varv)$ with $f_\alpha$ and $f_\beta$ the oscillator strengths of
both lines (actually doublets) and $\lambda_\alpha$ and $\lambda_\beta$ their
wavelengths. Furthermore, $f_{\rm c}(\varv )$ is the velocity-dependent covering
factor.

\begin{figure}[!tbp]
\resizebox{\hsize}{!}{\includegraphics[angle=-90]{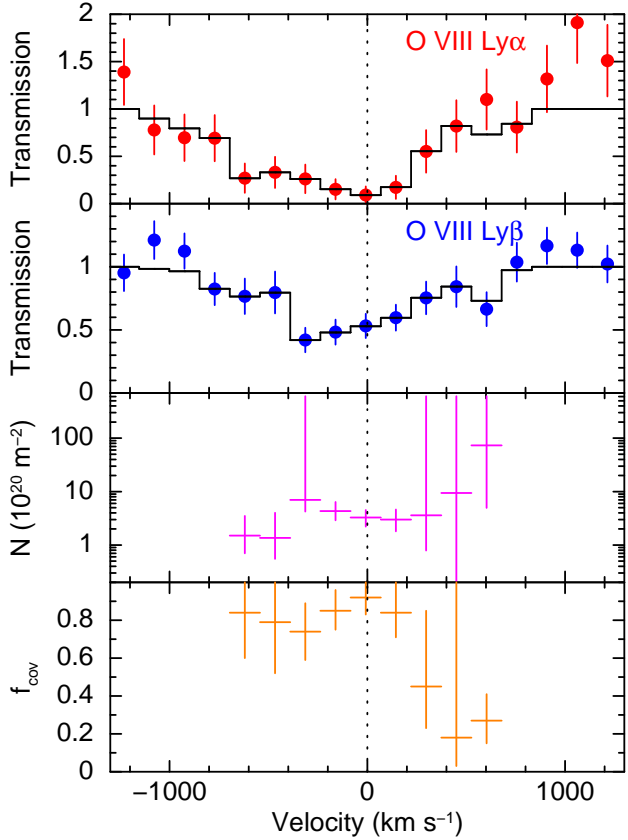}}
\caption{Velocity-dependent total \ion{O}{viii} column density $N$ and covering
factor $f_c$ (lower panels) deduced from fits (upper two panels) to the
\ion{O}{viii} Ly$\alpha$ and Ly$\beta$ lines.}
\label{fig:o8cov}
\end{figure}

We show our results in Fig.~\ref{fig:o8cov}. Only the portions of the profile
with well-determined covering factor or column density are shown. Between $-700$
and $+200$~km\,s$^{-1}$ the covering factor is fairly constant with a mean value
of $0.86\pm0.05$, because over this velocity range the Ly$\alpha$ line is deep
but not completely black. While the continuum near this line has about 15
counts/bin, the line core has 15 net counts distributed over 6 bins (the
subtracted background is negligible). 

The covering factor for the red wing of the line (between $+200$ and
$+700$~km\,s$^{-1}$) appears to be lower, on average about $0.31\pm 0.10$.

We have ignored here the contribution from any underlying broad emission line.
If we add the contribution from an emission line with a width (4200~km\,s$^{-1}$
FWHM) and strength (peak flux 6.7\% of the continuum) equal to that derived by
\citet{detmers2011} for the RGS-spectrum of Mrk~509, and assume that the
covering factor for the line is equal to that for the continuum, the average
covering factor in the region below and above $+200$~km\,s$^{-1}$ increases to
$0.87\pm 0.09$ and $0.32\pm 0.16$, respectively. This is only a modest
difference.

\section{Detailed spectral modelling of the outflow}\label{sect:outflow}

A detailed analysis of line profiles is only possible for the stronger lines
(see the previous section). Full spectral fitting is needed to make use of the
information contained in all absorption lines, including the weaker lines.  Such
an analysis was done using the RGS spectrum \citep{detmers2011}. Because the
HETGS has 3--6 times higher spectral resolution than the RGS, and the line
profiles of some lines (for instance \ion{O}{viii} 1s--2p) seem to indicate the
presence of outflow components with higher velocity than considered by
\citet{detmers2011}, we repeat their analysis but taking into account also the
HETGS data.

\subsection{Choice of velocity components}

The spectral resolution of the HETGS is significantly better than that of the
RGS, but still we lack the velocity resolution that is available in the UV.
Taking too many not fully independent velocity components in a spectral model
leads to strong correlations between these components and a poorly defined
solution. It is therefore necessary to limit the number of velocity components
taking into account the instrumental resolution. The FWHM of the MEG for the
strongest lines (\ion{O}{viii}) is about 300~km\,s$^{-1}$. For the HEG we get a
similar FWHM but now for the shorter wavelength lines like \ion{Ne}{x} and
\ion{Mg}{xii}. Ideally, a velocity bin size of about 1/3 FWHM would be
achievable, but given the limits imposed by the statistics of the spectrum
(relatively low effective area in the oxygen region), a minimum bin size of
about 150--200~km\,s$^{-1}$ seems appropriate.

The lowest velocity gas found by \citet{detmers2011} is component B1, with a
corrected velocity of $+80\pm30$~km\,s$^{-1}$. This component agrees fairly well
in velocity, column density and ionisation parameter with the UV troughs T5--T7
defined by \citet{arav2012}, with centroid velocities at $-15$, $+45$ and
$+125$~km\,s$^{-1}$, respectively. The column-density weighted average velocity
of these UV components is about $+$50~km\,s$^{-1}$, and we will adopt that
velocity as one of our basic components. From Fig.~\ref{fig:neo} we see that
there is some material at such velocities for lines with ionisation parameter
$\log\xi$ less than about 2.0 (\ion{O}{viii}).

The UV data show a lack of material with velocities between about $-200$ and
$-100$~km\,s$^{-1}$. The highest velocities measured in the UV are from the
troughs T1 ($-405$~km\,s$^{-1}$), T2 ($-310$~km\,s$^{-1}$), and T3
($-240$~km\,s$^{-1}$). From our individual line analysis, we find that the line
centroid of the lines from the ions that are dominated by ionisation component D
in the notation of \citet{detmers2011} (\ion{Si}{xiv}, \ion{Si}{xiii},
\ion{Mg}{xii} and \ion{Ne}{x}) have a weighted average outflow velocity of
$-227\pm24$~km\,s$^{-1}$, consistent with the velocity of UV trough T3. Hence,
we will also adopt a velocity of $-240$~km\,s$^{-1}$ as one of our basic
components.

The third velocity component then naturally corresponds to trough T1 at
$-405$~km\,s$^{-1}$. Given the resolution of the HETGS, we will not be able to
fully disentangle the troughs T1--T3, but by choosing T1 and T3 as our base
components we will be able to model the X-ray velocity structure with sufficient
accuracy. Any gas present at the velocities of T2 will be partly assigned to T1
and partly to T3.

Finally, \citet{detmers2011} found some evidence of higher velocity gas at
around $-700$~km\,s$^{-1}$, mainly in the \ion{Mg}{xi} and \ion{Fe}{xx} and
\ion{Fe}{xxi} ions. Our line profile for \ion{Mg}{xii} (Fig.~\ref{fig:simg})
seems to exclude such a component, although the broad blue wing of \ion{O}{viii}
might be consistent with such a component. Therefore we will include a fourth
velocity component labelled T0 at $-700$~km\,s$^{-1}$.

We summarise our adopted components in Table~\ref{tab:velcomp} below.

\begin{table}[!htbp]
\caption{Adopted velocity components for the analysis of the HETGS spectrum.} 
\centering                
\begin{tabular}{lcccc} 
\hline\hline
component & T0 & T1 & T3 & T5--T7 \\
\hline
$\varv$ (km\,s$^{-1}$) & $-700$ & $-405$ & $-240$ & $+50$ \\
\hline
\label{tab:velcomp}                         
\end{tabular}
\end{table}   

\subsection{Spectral modelling}\label{sect:model}

We have first made a simultaneous fit to the RGS and HETGS data. The continuum
model that we used is a small modification of the model used by
\citet{detmers2011}. While they used a pure spline for the full continuum of the
RGS spectrum, we use the sum of a power-law component and a spline. The
power-law gives a fair description of the high-energy part of the HETGS spectrum
(apart from Fe-K line features). While \citet{detmers2011} used the spline to
describe the full continuum (power-law plus soft excess), now it describes only
the soft excess component. For the RGS, the spline has 12 knots, logarithmically
spaced between 5--40~\AA, with the flux value at 5~\AA\ forced to be zero, such
that for $\lambda < 5$~\AA\ the continuum is given by the power-law. For the
HETGS spectrum, we use a linear wavelength grid for the spline, with a spacing
of 2~\AA\ between 6.5--26.5~\AA, and again with zero flux at $\lambda=6.5$~\AA.
In our spectral fitting, the splines representing the soft excess are allowed to
vary independently for both datasets to account for weak time variability and
cross-calibration effects, both of which are most important at lower energies
where both instruments overlap. We keep the photon index of the power-law frozen
to a value determined from a preliminary fit to the HETGS data. Because the RGS
does not cover the $\lambda<5$~\AA\ regime, the precise value of the photon
index is not important for RGS; any differences with the "true" underlying
power-law in the soft band are compensated for by the spline.

Furthermore, we add three broad Gaussian emission lines for the 1s--2p
transitions significantly detected in the RGS spectrum: \ion{O}{viii},
\ion{O}{vii} and \ion{Ne}{ix}, with their width frozen to the value obtained by
\citet{detmers2011} for the RGS spectrum (FWHM 4200~km\,s$^{-1}$). The line
fluxes are also frozen to the values obtained from the RGS spectrum.

The foreground Galactic absorption model is also the same as used by
\citet{detmers2011} and described in more detail by \citet{pinto2012}.

We next add photo-ionised absorption components (the \textsl{xabs} model of
SPEX). This model describes the transmission of a thin layer of photo-ionised
gas.

We first start testing how many absorption components are needed. For that
purpose, we incorporate a total number of $7\times 4 =28$ \textsl{xabs}
components in our model. For each of the 4 velocity components, we use 7
different ionisation components, with values of $\log\xi$ frozen to 0, 0.5, 1,
1.5, 2, 2.5 and 3, respectively. For this spacing of ionisation parameter, any
gas with ionisation parameter in the range of $\log\xi$ between 0--3 will give a
detectable signal for at least 1 component. In addition, as discussed before our
spacing of the velocity components is sufficient to include all gas. The
velocity broadening was fixed to $\sigma_{\rm v} = 100$~km\,s$^{-1}$ for each
component. Only the column densities of all 28 components are free parameters.
This is sufficient for the present purpose, namely to detect all relevant
spectral components.

\begin{figure}[!tbp]
\resizebox{\hsize}{!}{\includegraphics[angle=-90]{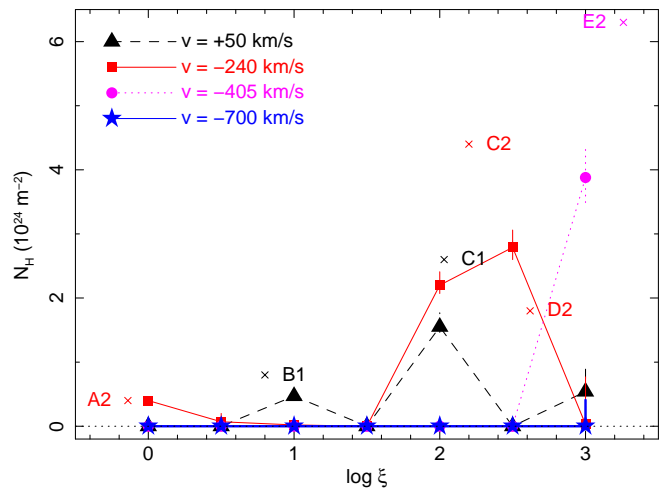}}
\caption{Best-fit column densities for a model with 28 absorption components
applied to the joint RGS and HETGS spectra. Crosses with labels indicate the
positions of the components found in the analysis of the RGS spectrum by
\citet{detmers2011}. }
\label{fig:curpar0}
\end{figure}

Fig.~\ref{fig:curpar0} shows the result of a joint fit to the RGS and HETGS
spectrum. At the lowest velocity ($+50$~km\,s$^{-1}$) we find two components,
close to components B1 and C1 defined by \citet{detmers2011}. For gas at
$-240$~km\,s$^{-1}$ a blend of the components C2 and D2 is recovered: it should
be noted that in the analysis of \citet{detmers2011} the ionisation parameters
and outflow velocities were free parameters, giving more flexibility than in the
present analysis with fixed ionisation parameters. Component A2 is also
recovered at this velocity. Finally, at $-405$~km\,s$^{-1}$ we find gas at
$\log\xi=3$, corresponding to component E2. For this highest ionised component,
there is a strong correlation between column density and ionisation parameter
such that the 1$\sigma$ lower limit for component E2 (not shown in the figure)
is closer to our component at $\log\xi=3$ both in ionisation parameter and
column density.

It thus appears that there is little room for more absorption components than
the ones already found by \citet{detmers2011}. However, to test this in more
detail, we have also investigated another model.

We use a model with 8 \textsl{xabs} components, but contrary to our analysis
above we allow three main parameters ($\log\xi$, $N_{\rm H}$, and the velocity
dispersion $\sigma_{\rm v}$) to be free. The outflow velocities are frozen to
one of the values from Table~\ref{tab:velcomp}. For six of these components, we
use initial conditions close to the values for components A2, B1, C1, C2, D2 and
E2 of \citet{detmers2011}. To test whether there is gas present at $\log\xi>3$
we added a seventh component with a velocity of $-405$~km\,s$^{-1}$ and a high
ionisation parameter. The HETGS, because of its wavelength range, is well suited
to detect highly ionised absorption. For component 8 we use an outflow velocity
of $-700$~km\,s$^{-1}$ with an initial ionisation parameter in the range where
\ion{O}{viii} has its peak concentration, in order to see if we can confirm the
blue wing of the \ion{O}{viii} line (Fig.~\ref{fig:neo}). We execute the
spectral fits three times: for the RGS only, HETGS only and for the combined RGS
and HETGS spectrum.  

\begin{table*}[!htbp]
\caption{Parameters of the spectral fits with multiple ionisation components} 
\centering                
\begin{tabular}{@{}l@{}l|cccccccc@{}} 
\hline\hline
Conf\tablefootmark{a} & & B1 & C1 & A2 & C2 & D2 & E2 & -- & -- \\ 
\hline 
 & $v$\tablefootmark{b}         &  $+50$       & $+50$         &$-240$         &$-240$        &$-240$          &$-405$        &$-405$        & $-700$         \\
\hline 
R & $\log\xi$\tablefootmark{c}  &$0.96\pm 0.08$&$ 1.96\pm 0.02$&$-0.07\pm 0.09$&$2.11\pm 0.02$&$2.65\pm 0.07  $&$2.93\pm 0.08$&$3.54\pm 0.09$&$3.30\pm  0.19 $\\
RH& $\log\xi$\tablefootmark{c}  &$0.98\pm 0.07$&$ 2.02\pm 0.04$&$-0.04\pm 0.09$&$2.07\pm 0.02$&$2.66\pm 0.04  $&$2.98\pm 0.22$&$3.64\pm 0.08$&$3.29\pm  0.09 $\\ 
H & $\log\xi$\tablefootmark{c}  &$1.09\pm 0.32$&$ 1.95\pm 0.11$&$ 0.23\pm 0.22$&$1.91\pm 0.09$&$2.64\pm 0.07  $&$3.27\pm 0.24$&$3.85\pm 0.33$&$3.24\pm  0.25 $\\ 
\hline 
R & $N_{\rm H}$\tablefootmark{d}&$0.53\pm 0.07$&$ 1.35\pm 0.19$&$ 0.53\pm 0.05$&$ 4.2\pm  0.4$&$ 2.0\pm  1.0  $&$ 3.0\pm  0.8$&$   6  (<14) $&$ 1.1   (<2.7) $\\ 
RH& $N_{\rm H}$\tablefootmark{d}&$0.56\pm 0.06$&$ 1.57\pm 0.19$&$ 0.54\pm 0.05$&$ 3.3\pm  0.3$&$ 4.0\pm  0.6  $&$ 1.8\pm  0.9$&$  10  (<22) $&$ 1.7\pm 0.9   $\\ 
H & $N_{\rm H}$\tablefootmark{d}&$0.70\pm 0.33$&$ 1.57\pm 0.44$&$ 0.56\pm 0.24$&$ 1.5\pm  0.5$&$ 5.9\pm  1.0  $&$ 1.7  (<4.3)$&$  19   (<49)$&$ 1.9   (<4.6) $\\ 
\hline 
R & $\sigma_\varv$\tablefootmark{e}&$ 83\pm 13$&$  113\pm   15$&$   52\pm   12$&$  75\pm   10$&$  39\pm   24  $&$  90\pm   60$&$  40  (<260)$&$     -	     $\\  
RH& $\sigma_\varv$\tablefootmark{e}&$ 82\pm 13$&$  111\pm   14$&$   52\pm   11$&$  86\pm   11$&$  39\pm   15  $&$ 130\pm  110$&$   <100     $&$340(-240,+350)$\\ 
H & $\sigma_\varv$\tablefootmark{e}&$100\pm 90$&$  160\pm   70$&$   90\pm   50$&$  60	(>30)$&$ 170(-130,+40)$&$  90  (<490)$&$   <800     $&$   <2200      $\\
\hline 
\label{tab:fits}                         
\end{tabular}
\tablefoot{
\tablefoottext{a}{Configuration. R: RGS only; RH: RGS \& HETGS; H: HETGS only}
\tablefoottext{b}{Outflow velocity in km\,s$^{-1}$}
\tablefoottext{c}{Ionisation parameter in $10^{-9}$~W\,m as used throughout
this paper}
\tablefoottext{d}{Total hydrogen column density in $10^{24}$~m$^{-2}$}
\tablefoottext{e}{Velocity dispersion in km\,s$^{-1}$}
}\end{table*}   

The best-fit parameters are shown in Table~\ref{tab:fits}. We first consider the
already known components B1--E2. In general, the statistical uncertainties on
the HETGS spectrum are somewhat larger than on the RGS spectrum. This is caused
by the factor of two difference in exposure time and the larger effective area
at longer wavelengths, both in favour of the RGS. Interestingly, the only
significant variability appears to occur for component C2 (4.2$\sigma$
significance decrease) and D2, with a tripling of the column density
(2.8$\sigma$ significance).

The additional two components are poorly constrained. The column density
of the component at $-405$~km\,s$^{-1}$  must be regarded as an upper limit.
The component at $-700$~km\,s$^{-1}$ also converged to a high-ionisation value,
with low significance and poor constraints. Again, the column density is merely
an upper limit.

\subsection{Abundances}

\begin{table}[!htbp]
\caption{Best-fit abundances in the outflow of Mrk~509 relative to oxygen and
in proto-solar units.} 
\centering                
\begin{tabular}{lccc} 
\hline\hline
Element & RGS\tablefootmark{a} & HETGS\tablefootmark{b} 
  & RGS+HETGS\tablefootmark{b}\\
\hline
C/O  & $1.19\pm0.08$ &  --           & $1.17\pm0.07$\\
N/O  & $0.98\pm0.08$ &  --           & $1.04\pm0.07$\\
Ne/O & $1.11\pm0.10$ & $1.11\pm0.10$ & $1.20\pm0.07$\\
Mg/O & $0.68\pm0.16$ & $1.09\pm0.13$ & $0.92\pm0.11$\\
Si/O & $1.3 \pm0.6 $ & $1.09\pm0.15$ & $0.98\pm0.14$\\
S/O  & $0.57\pm0.14$ & $0.5 \pm0.4 $ & $0.80\pm0.17$\\
Ca/O & $0.89\pm0.25$ &      --       & $1.30\pm0.34$\\
Fe/O & $0.85\pm0.06$ & $0.87\pm0.09$ & $0.89\pm0.03$\\
\hline
\label{tab:abu}                         
\end{tabular}
\tablefoot{
\tablefoottext{a}{From \citet{steenbrugge2011}; all abundances are from their
analysis of the RGS data, except Si that was derived from the LETGS data.}
\tablefoottext{b}{Present work.}
}\end{table}   

Using our model with the six significant \textsl{xabs} components presented in
the previous section we have also determined the abundances in the outflow by
making them free parameters. As discussed in \citet{steenbrugge2011}, we keep
the oxygen abundance frozen to the proto-solar value \citep{lodders2009},
because the X-ray spectrum contains no hydrogen or helium lines. Thus, all
abundances are relative to oxygen. We determined the abundances using both the
combined RGS and HETGS spectrum as well as using only the HETGS spectrum. We
show our results in Table~\ref{tab:abu}. The abundances of the combined fit are
not always strictly between those derived from the individual RGS and HETGS
fits. This is caused by the different bandpass of both instruments and the
slightly different spectral models, but in general the abundances agree within
their nominal statistical uncertainties.

\subsection{The 2--6~keV band}

Our HETGS spectrum has the best combination of sensitivity with spectral
energy resolution over the 2--6~keV band. We have searched for the presence of
absorption and emission lines in this band, but there are no features stronger
than our detection limit of about 3~m\AA\ equivalent width in this band. This is
consistent with the prediction from our model for the outflow presented before
in this section. The same model also predicts that the lines from innershell
absorption lines in the Fe-K band are below the detection limit of the HETGS
data. Limits for ultra-fast outflows in this band are discussed in
Section~\ref{subsect:ufo}.

\section{The Fe-K complex}\label{sect:fek}

\subsection{General considerations}

Our present HETGS spectrum is the best-exposed high-resolution spectrum of
Mrk~509 in the Fe-K band. \citet{yaqoob2004} detected a narrow iron line in a
shorter, 59~ks HETGS spectrum taken in 2001. The analysis of the XMM-Newton pn
spectrum \citep{ponti2013} shows that the neutral Fe-K emission line has two
components: a narrow component that remains constant in flux over time, and a
broader component with a constant equivalent width from days to years
timescales. In addition, the XMM-Newton spectrum shows excess flux between
6.7--7.0 keV that can be modelled well with two different models: a weak
relativistically broadened emission line or weak highly-ionised line emission
from H-like or He-like iron, possibly produced by scattering from distant
material.

\begin{figure}[!tbp]
\resizebox{\hsize}{!}{\includegraphics[angle=-90]{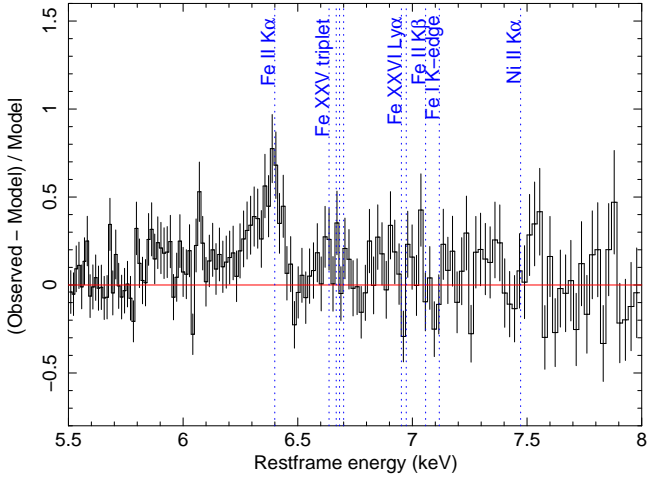}}
\caption{HEG spectrum near the Fe-K emission line. Shown are the residuals
relative to a power-law continuum. The rest frame energies of some important
transitions are indicated by dotted lines (the \ion{Fe}{xxv} Ly$\alpha$ line
consist of two components, and the \ion{Fe}{xxv} triplet consists of four
components, from left to right the forbidden line, the intercombination lines
and the resonance line).}
\label{fig:fekplot}
\end{figure}

\begin{figure}[!tbp]
\resizebox{\hsize}{!}{\includegraphics[angle=-90]{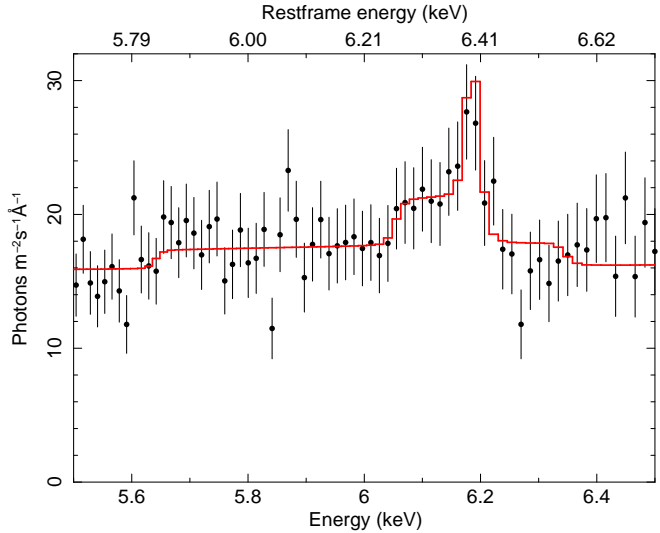}}
\caption{HEG spectrum in 2012 near the Fe-K emission line (data points with
error bars). The model (solid line) consists of a power-law, narrow emission
line and two flat plateaus.}
\label{fig:compton}
\end{figure}

Our HEG spectrum near the Fe-K complex is shown in Fig.~\ref{fig:fekplot}. The
spectrum is shown relative to the power-law continuum obtained from the model
fit discussed below (Fig.~\ref{fig:compton}). The most obvious feature in the
spectrum is the narrow emission line at 6.4 keV rest frame energy which shows a
red shoulder extending down to 6.25~keV, and perhaps even down to about 5.8~keV.
We treat these features in more detail in Sect.~\ref{sect:narrow} below. Because
of the statistical uncertainties of the spectrum, we cannot prove the presence
of other spectral features at higher energies, although, for instance, the
residuals in Fig.~\ref{fig:fekplot} near the \ion{Fe}{xxv} and \ion{Fe}{xxvi}
1s--2p transitions are consistent with the predictions from the XMM-Newton model
of \citet{ponti2013} (equivalent width about 4~eV for each line).

\citet{ponti2013} described the 6.4~keV line by the sum of a narrow and broad 
($\sigma=0.22$~keV) Gaussian line. Our data are still consistent with such a
model and the strengths of the lines derived by \citet{ponti2013}, but this
model cannot reproduce the strong red wing of the narrow line, unless the broad
line would be four times stronger and would lack a blue wing.
Therefore we model this spectral region with a different model as outlined in
the next section.

\subsection{Analysis of the Fe-K emission line\label{sect:narrow}}

The spectrum near the narrow line at 6.4~keV rest frame energy is shown in
Fig.~\ref{fig:compton}. In addition to the narrow line, there is a broad red
wing extending down to 6.25~keV, and perhaps even down to about 5.8~keV.  

We have fitted the continuum first using the continuum model described in
Sect.~\ref{sect:model}, i.e. a high-energy powerlaw and a spline at low energies
representing the soft excess. We added the outflow as modelled by the sum of
\textsl{xabs} components but with their parameters frozen to the values found in
Sect.~\ref{sect:model}. Only the parameters of the power-law and the spline were
allowed to vary.

We then added a narrow iron line at 6.40~keV to the model (assumed to be a
$\delta$-line). The red wing of the line extending down to $\sim$6.25~keV could
suggest perhaps the presence of a Compton shoulder. In order to test this, we
have added a flat plateau to the model. It consists of a $\delta$-line with
centre 6.336~keV, smeared with a flat profile (the \textsl{vblo} component of
SPEX) with a full-width of 0.15~keV. This mimics to lowest order the Compton
shoulder profiles as presented e.g. by \citet{matt2002}. The even broader red
wing is modelled here with a similar flat profile with centre 6.2~keV and
half-width of 0.73~keV.

We obtain a best fit normalisation of the narrow line of $0.09\pm
0.03$~ph\,m$^{-2}$\,s$^{-1}$ or an equivalent width of 16$\pm$5~eV. For the
narrow plateau we obtain a flux of $0.19\pm 0.06$~ph\,m$^{-2}$\,s$^{-1}$ or an
equivalent width of 33$\pm$10~eV. The broadest plateau has a flux of $0.41\pm
0.10$~ph\,m$^{-2}$\,s$^{-1}$. Without the broadest plateau in the model, the
narrow line would have a slightly higher flux of 0.10~ph\,m$^{-2}$\,s$^{-1}$ and
its red wing a flux of 0.28~ph\,m$^{-2}$\,s$^{-1}$. See Fig.~\ref{fig:compton}
for this empirical fit.

As an alternative model, we have used the same model but with the two plateau
components replaced by a relativistically blurred emission line. For that
purpose we use the \textsl{refl} model of SPEX, which was provided by Piotr
Zycki. It calculates the reflected continuum plus the corresponding Fe-K line
from a constant density X-ray illuminated atmosphere. It computes the
Compton-reflected continuum \citep[cf.][]{magdziarz1995} and the iron K$\alpha$
line \citep[cf.][]{zycki1994}, as described in \citet{zycki1999}. In addition it
can be convolved with a relativistic discline model (for Schwarzschild
geometry).

\begin{figure}[!tbp]
\resizebox{\hsize}{!}{\includegraphics[angle=-90]{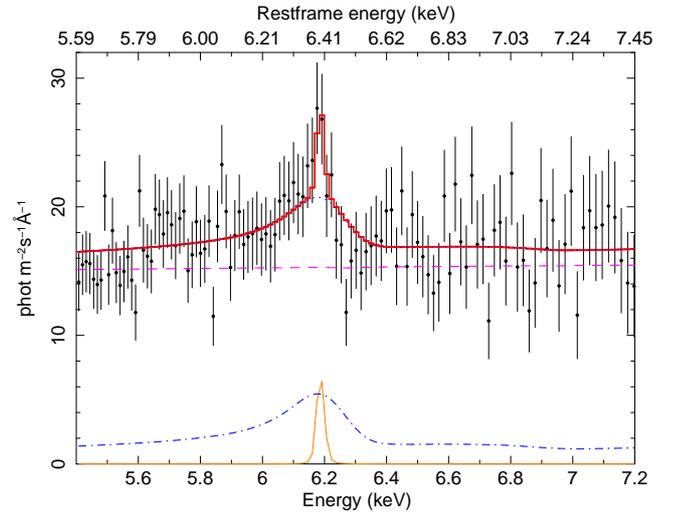}}
\caption{HEG spectrum near the Fe-K emission line (data points with error bars)
fitted to a model with a narrow emission line and a relativistically blurred
cold reflection component.
The lines are from bottom to top: contribution from the narrow emission line
(solid), reflection component (dot-dashed), powerlaw component (dashed),
sum of power law and reflection (dotted) and total model (solid histogram).}
\label{fig:relline}
\end{figure}

\begin{table}[!htbp]
\caption{Model for the Fe-K emission line region.} 
\centering                
\begin{tabular}{ll} 
\hline\hline
Parameter & value \\
\hline
$\Gamma$\tablefootmark{a} & $1.924\pm0.010$ \\
$R$\tablefootmark{b} & $0.64\pm0.11$ \\
$q$\tablefootmark{c} & $-2.23\pm0.20$ \\
$\cos i$\tablefootmark{d} & $0.95$ ($>0.944$)\\
$F_{\rm{refl}}$\tablefootmark{e} & 3.9~ph\,m$^{-2}$\,s$^{-1}$\\
$F_{\rm{line}}$\tablefootmark{f} & $0.09\pm 0.03$~ph\,m$^{-2}$\,s$^{-1}$\\
$\sigma_{\rm{line}}$\tablefootmark{g} & $<2900$~km\,s$^{-1}$\\
\hline
\label{tab:relline}                         
\end{tabular}
\tablefoot{
\tablefoottext{a}{Photon index of the primary continuum.}
\tablefoottext{b}{Scale $R$ for reflection. For an isotropic source above an
infinite disc $R = 1$. This value corresponds to seeing equal contributions
from the reflected and direct spectra.}
\tablefoottext{c}{Slope of the emissivity $\epsilon$ of the disc, power-law
$\epsilon\sim r^q$.} 
\tablefoottext{d}{Cosine of inclination angle $i$ of the line of sight to the
disc. The fit reached the maximum (hard) value of the model.}
\tablefoottext{e}{Total flux of the reflection component. Not a free parameter
but derived from the fit.}
\tablefoottext{f}{Flux of the narrow emission line at 6.40~keV (energy fixed).}
\tablefoottext{g}{Gaussian width of the narrow emission line. We find an upper
limit. For practical purposes, we keep its value to 2000~km\,s$^{-1}$.}
}\end{table}   

This model gives an acceptable solution, but because the corresponding line is
not very strong, we have kept a few parameters fixed. We kept the inner- and
outer disc radius frozen to 6 and $10^4$ gravitational radii, the metallicity
and iron abundance to solar values and the ionisation parameter to zero (i.e.
cold reflection). We show our solution in Table~\ref{tab:relline} and
Fig.~\ref{fig:relline}.

\section{Ultra-fast outflows}\label{sect:ufos}

\subsection{Fe-K region}\label{subsect:fekregion}

Ultra-fast outflows \citep[e.g.][]{tombesi2010} have been previously reported
for Mrk~509. \citet{dadina2005} report a redshifted system ($+0.21$c if
interpreted as \ion{Fe}{xxvi}) in two out of five time segments of one of the
six BeppoSAX observations taken between 1998--2001, as well as a blue-shifted
system at $-0.16$c in one of these time segments. In addition, they report an
outflow system at $-0.19$c in the XMM-Newton observation taken in 2000.
\citet{cappi2009} reported three systems in three different XMM-Newton
observations taken in 2000, 2005 and 2006. They have equivalent widths of $32\pm
12$, $20\pm 6$ and $20\pm 7$~eV at outflow velocities of $-0.172$c, $-0.141$c
and $-0.196$c, respectively, when identified with \ion{Fe}{xxvi} 1s--2p. In the
stacked XMM-Newton data taken before 2009, \citet{ponti2009} found evidence of
an additional system with an outflow velocity of $-0.048$c and with equivalent
width of $13(-4,+2)$~eV (the error bars from both papers have been converted to
68\% here). The lack of a similar absorption line during the non-simultaneous
Suzaku campaign was interpreted by \citet{ponti2009} as being due to intrinsic
variability in the absorber properties. Moreover, the stacked XMM-Newton pn
spectrum from the deep exposure taken in 2009 does not show any significant
($>3\sigma$) evidence of ultra-fast outflows in the Fe-K band \citep{ponti2013}.

\begin{figure}[!tbp]
\resizebox{\hsize}{!}{\includegraphics[angle=-90]{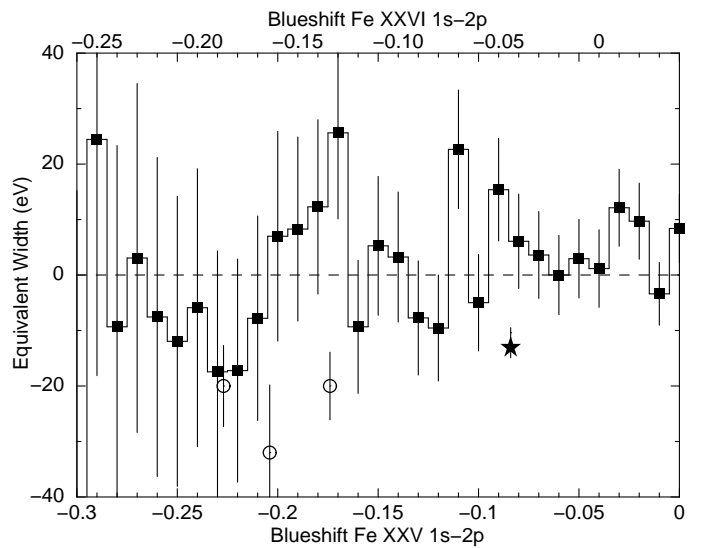}}
\caption{Best-fit equivalent width (EW) of an iron 1s--2p absorption line for
the present spectrum taken with the HEG. The lower x-axis corresponds to the
blueshift of an \ion{Fe}{xxv} line, the upper x-axis to that of an
\ion{Fe}{xxvi} line. In this plot positive EW corresponds to emission, negative
EW to absorption. Filled squares: present HEG data; open circles: the
three components reported by \citet{cappi2009}; star: the component
discussed by \citet{ponti2009}.}
\label{fig:ufo}
\end{figure}

We have searched for the presence of absorption lines from highly ionised iron
ions in the present HETGS data by adding to our best-fit continuum model
discussed before absorption from \ion{Fe}{xxv} or \ion{Fe}{xxvi} at various
redshifts. Because with the standard extraction that we used for our HEG spectra
the spectrum for energies above $\sim$7.6~keV starts losing counts, that are not
taken into account in the fluxed spectrum, we replaced the fluxed spectrum above
that energy with the fluxed spectrum taken from the Chandra TGCAT
archive\footnote{See tgcat.mit.edu}, that explicitly uses a narrower
extraction region to improve the spectrum in the 1.5--1.8~\AA\ range. We
verified that at $E<7.6$~keV the fluxed spectrum is exactly the same using both
extraction methods.

We have frozen the velocity broadening to 1000~km\,s$^{-1}$ for this exercise,
to avoid line saturation while still being moderately narrow compared to the
instrumental resolution of the HEG grating at these energies. The resulting
equivalent widths are shown in Fig.~\ref{fig:ufo}. We find no significant
absorption lines, and the numbers and their error bars can be regarded as upper
limits. 

A typical upper limit of 20~eV equivalent width corresponds to a column density
of $2.0\times 10^{21}$ or $4.7\times 10^{21}$~m$^{-2}$ for \ion{Fe}{xxv} and
\ion{Fe}{xxvi}, respectively; for proto-solar abundances this corresponds to a
minimum hydrogen column density of $1.2\times 10^{26}$ and $3.1\times
10^{26}$~m$^{-2}$, at the respective ionisation parameters where the peak
concentration occurs ($\log\xi$ values of 3.7 and 4.1, respectively).

\subsection{Other possible ultra-fast outflow signatures\label{subsect:ufo}}

The Chandra LETGS spectrum of Mrk~509 taken in 2009 showed a few weak features
in the 5--9~\AA\ band that might be associated with highly ionised gas
outflowing at about $-14\,500$~km\,s$^{-1}$ \citep{ebrero2011}. The strongest
features would be the 1s--2p transitions of \ion{Mg}{xii}, \ion{Si}{xiv} and
\ion{S}{xv}. While such a component is only significant at the 90\% confidence 
level, its outflow velocity coincides with the velocity of one of the highly
ionised outflowing components seen in the iron line as reported by
\citet{ponti2009}. This provides sufficient motivation to inspect our HETGS
spectrum carefully.

\begin{table}[!htbp]
\caption{Best-fit column densities in $10^{20}$~m$^{-2}$ for highly ionised ions
at $-14\,500$~km\,s$^{-1}$.} 
\centering                
\begin{tabular}{lcccc} 
\hline\hline
Ion & HETGS\tablefootmark{a} & LETGS\tablefootmark{b} \\
\hline
\ion{Mg}{xii} & 0.3$\pm$0.3 & 4$\pm$8 \\
\ion{Si}{xiv} & 0.0$\pm$0.5 & 5$\pm$12 \\
\ion{S}{xv}   & 0.3$\pm$0.7 & $<700$ \\
\hline
\label{tab:simg_ufo}                         
\end{tabular}
\tablefoot{
\tablefoottext{a}{Present HETGS spectrum, 280~ks, taken in 2012.}
\tablefoottext{b}{LETGS spectrum, 170~ks, taken in 2009 \citep{ebrero2011}.}
}\end{table}   

We have added, similar to \citet{ebrero2011}, a \textsl{slab} component to our
best-fit model of the HETGS spectrum, with velocity dispersion frozen to
100~km\,s$^{-1}$ and outflow velocity to $-14\,500$~km\,s$^{-1}$. The
\textsl{slab} model accounts for the transmission of a layer of material
composed of ions with adjustable ionic column densities. The best-fit column
densities are reported in Table~\ref{tab:simg_ufo}. We find no evidence of such
a component. Our upper limits are more than an order of magnitude lower than
those reported by \citet{ebrero2011}. For proto-solar abundances, they
correspond to an equivalent hydrogen column density of $<3\times
10^{24}$~m$^{-2}$ at a typical ionisation parameter of $\log\xi = 3$.

\section{Discussion}\label{sect:discussion}

\subsection{Wavelength scale}

In \citet{kaastra2011b} we paid a lot of attention to obtaining the proper
wavelength scale for the RGS spectra, because of small systematic uncertainties
of the order of 7~m\AA. In the present work we have found that the velocities
based on the RGS spectra reported by \citet{detmers2011} need to be increased by
$+55\pm 21$~km\,s$^{-1}$ to bring them into agreement with the HETGS scale.  The
absolute wavelength accuracy of the MEG grating as reported in the Chandra
observatory guide is 11~m\AA, corresponding to about 150~km\,s$^{-1}$ for the
oxygen lines. However, the relative uncertainty is only 2~m\AA\ or
30~km\,s$^{-1}$ within and between observations. We think the latter may better
represent the uncertainty for our observations. The correction of
$+$55~km\,s$^{-1}$ corresponds to +3.7~m\AA\ at a typical wavelength of 20~\AA;
adjusting the RGS wavelength scale indicators in Table~11 of
\citet{kaastra2011b} with this number brings them all in good agreement, where
we need to assume that most hot gas in our Galaxy in the direction towards
Mrk~509 is at low velocity \citep[cf.][]{pinto2012}. The only exception is the
1s--2p line of \ion{O}{i} from our Galaxy, that is off by about 10~m\AA. Perhaps
this line suffers from more blending by weak lines from the outflow of Mrk~509
than we have accounted for.

\subsection{Gas outflowing at $-700$~km\,s$^{-1}$ or higher}

\begin{figure}[!tbp]
\resizebox{\hsize}{!}{\includegraphics[angle=-90]{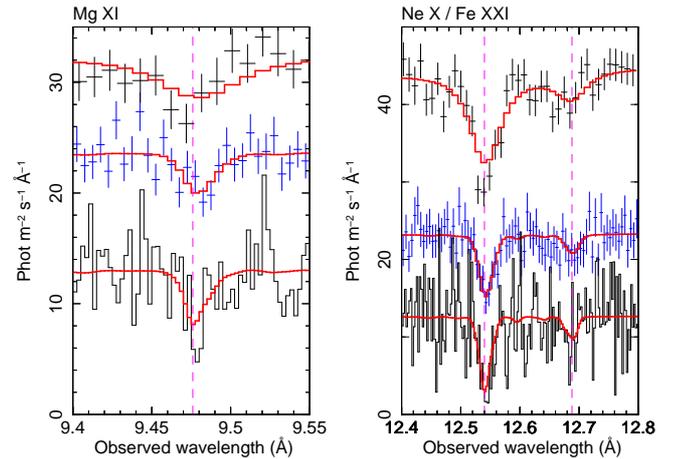}}
\caption{Observed spectra near the strongest \ion{Mg}{xi} (left panel) and
\ion{Ne}{x} (12.54~\AA) and \ion{Fe}{xxi} (12.69~\AA) transitions (right
panel).  Spectra from top to bottom: RGS, MEG and HEG. For clarity of display we
have omitted error bars on the HEG spectra; all spectra have an arbitrary offset
in the y-direction, and for better visibility of the weak lines the RGS spectra
have in addition been multiplied by a factor of 5 in both panels. The model
spectra (thick solid histograms) represent the best-fit HETGS spectrum; for the
RGS this model has been simply folded through the RGS data. }
\label{fig:highvel}
\end{figure}

One of the motivations for our present observation was to confirm the presence
of the tentative velocity component at $-715\pm 109$~km\,s$^{-1}$ in the
RGS spectrum \citep{detmers2011}. That component was visible in the strongest
transitions of \ion{Mg}{xi}, \ion{Fe}{xx} and \ion{Fe}{xxi}. The HETGS line
profiles of \ion{Mg}{xi}, \ion{Mg}{xii} and other highly ionised ions do not
show such a component (Fig.~\ref{fig:simg}), and this is confirmed by our
spectral fits that tend to give only upper limits for column densities at
such velocities. 

Is the high-velocity component found by \citet{detmers2011} real? To address
this question, we show the RGS and HETGS spectra near the strongest transitions
from \ion{Mg}{xi} and \ion{Fe}{xxi} (Fig.~\ref{fig:highvel}). Compared to the
best-fit HETGS model, the RGS spectrum shows two data bins, each about 1$\sigma$
below the model, at the blue side from the centre of the line. In a fit to RGS
data alone, these tend to make the outflow velocity higher indeed, but the
excess is not very significant compared to the HETGS model. For \ion{Fe}{xxi} a
roughly similar situation occurs, and combining this weak blue excess absorption
for both lines creates the $-715$~km\,s$^{-1}$ component. Our work shows that
such a component is not present in the 2012 HETGS spectrum, but of course we
cannot exclude that such a somewhat weaker, but transient component was present
in the 2009 RGS data. 

Another indicator for possible high-velocity gas is the \ion{O}{viii} 1s--2p
line profile in the MEG spectrum, that appears to indicate a blue tail extending
out to about $-600$~km\,s$^{-1}$ or perhaps even beyond $-1000$~km\,s$^{-1}$.
Again, a global spectral fit does not confirm this, although the detailed line
profile  (Fig.~\ref{fig:o8cov}) does not exclude the presence of a small amount
of \ion{O}{viii} ions around $-600$~km\,s$^{-1}$.

\subsection{Covering factor of the outflow}

In Sect.~\ref{sect:o8prof} we have investigated the line profiles of the
\ion{O}{viii} Lyman series. According to the modelling of \citet{detmers2011},
this ion is produced for about 75\% in the ionisation components C, and 16\% and
8\% in the ionisation components D and E, respectively.

We first consider the outflow components around $-240$ (C2 and D2) and
$-405$~km\,s$^{-1}$ (E2). Our analysis shows that, on average, there is some
evidence that the covering factor is less than unity. At
$\varv<-100$~km\,s$^{-1}$, the average covering factor is $0.81\pm 0.08$. We
note that also the UV \ion{C}{iv} and \ion{N}{v} lines in the same velocity
range show some evidence of covering factors less than unity
\citep{kriss2011,arav2012}, but the detailed covering factor profile is hard to
determine for \ion{O}{viii}.

A lower than unity covering factor might be caused by either a patchy outflow,
or by contributions to the continuum emission outside the region covered by the
outflow. We first consider this second possibility. Because components C and D
of the outflow have distances of $>70$~pc and $5-33$~pc, respectively
\citep{kaastra2012}, and the inclination angle is likely to be in the range of
20--40 degrees, the extent of the outflow perpendicular to the line of sight
will be of the order of 2--40~pc or more. Thus, if the outflow covers the
continuum source only partially, there should be a significant (of the order of
10\%) continuum X-ray flux contribution at distances of the order of 2--40~pc
from the nucleus. The torus region would be a likely candidate, but in order to
get a strong reflected X-ray flux from this region one would need a high
ionisation parameter in addition to a sizeable sustained solid angle with
respect to the nucleus. For such a high ionisation parameter significant atomic
edges should be present in the reflected spectrum, and these cannot be
relativistically blurred at such large distances from the nucleus. Our RGS
spectrum also shows no evidence of strong edges.

Thus, a somewhat patchy outflow may be a better explanation for the slightly
reduced covering factor, but we must stress here that the significance of
$f_{\rm c}<1$ in this velocity range is $\sim 2.4\sigma$, so there is a small
probability that the covering factor is 1 and that the column density is on the
lower side in order to make the \ion{O}{viii} line core not completely black.

We next consider the velocity range between $-100$ and $+200$~km\,s$^{-1}$,
around the low velocity components. Here the average covering factor for
\ion{O}{viii} is $0.90\pm 0.08$, hence it is hard to draw firm conclusions. At
even higher redshifts, for velocities $>200$~km\,s$^{-1}$ the covering factor
appears to be significantly smaller, about 0.3. However, in the UV lines there
is no evidence of gas at these velocities, and most of the low covering factor
is caused by the single bin at $+600$~km\,s$^{-1}$. At these velocities there is
some weak blending with \ion{Fe}{xii} transitions; this ion peaks at about the
same ionisation parameter as \ion{O}{viii} and according to the model of
\citet{detmers2011} it can contribute a few percent to the transmission of the
\ion{O}{viii} 1s--3p line around 16.03~\AA\ rest frame energy.

\subsection{Variability}

Our analysis in Sect.~\ref{sect:outflow} shows that, within the statistical
uncertainties, there have been no changes in the derived column densities of
most regular outflow components. The only exception appear to be components C2
and D2, both at $-240$~km\,s$^{-1}$ and with $\log\xi$ at 2.07 and 2.66,
respectively. Component D2 is the only X-ray component that showed significant
long-term variability \citep{kaastra2012}, and at a distance between 5--33~pc it
is also the component nearest to the central black hole. In the three years
between 2009 and 2012 it may well be that fresh material has entered the outflow
in this component, if it is near its lower limit distance of 15 light years
(5~pc). For component C2 such an explanation is unlikely, given its distance of
$>70$~pc. We note that components C2 and D2 are close to unstable branches of
the cooling curve \citep{detmers2011}; thus, depending on the luminosity
history, gas may have moved from one ionisation parameter to another. We note
that the loss of column density for component C2 is about the gain in column
density for component D2. Lacking more regular monitoring of Mrk~509, however,
we cannot test such a scenario.

\subsection{Abundances}

Our best-fit abundances do not differ much from those reported by
\citet{steenbrugge2011} based on the RGS data. We refer to that paper for a full
discussion on the abundances. In general, inclusion of the HETGS data improves
the accuracy, in particular for Ne, Mg, Si and Fe. The Mg and S abundances are
somewhat closer to the proto-solar values by including the HETGS data.

\subsection{Fe-K emission line}

The spectrum clearly shows a narrow emission line component. Its flux is
consistent with the values found by \citet{yaqoob2004} and \citet{ponti2013},
and its steadiness in flux over the years points to an origin far away from the
nucleus.

The narrow iron line at 6.4~keV appears to have an extended red wing extending
down to about 6.25~keV (Fig.~\ref{fig:compton}). This looks quite similar to a
Compton shoulder, and for that reason we have attempted a fit with a flat profile
(Sect.~\ref{sect:fek}). This gives a remarkable good fit to the observed
profile, but the shoulder has about two times more photons than the narrow line.
All models calculated by \citet{matt2002} for Compton shoulders have a shoulder
to line ratio $f$ less than 0.4 for transmission models and $f<0.2$ for
reflection models. We can clearly exclude a transmission model here, because
that would give significant low-energy absorption that is not observed. Hence,
there is a mismatch of a factor of 10 in the observed value for $f$ compared to
the models. This is even stronger if we consider that the equivalent width of
the narrow emission line is small (16~eV), and for such a small value the models
of \citet{matt2002} even produce a lower value $f<0.10$. Despite the good fit,
it thus appears hard to explain the red wing with a Compton shoulder, and we
therefore have to seek another explanation for this red wing.

The full width of the red wing is about 7000~km\,s$^{-1}$. This is smaller, but
of a similar order to the line width detected by \citet{ponti2013}. If we would
assume that this emission stems from an X-ray broad line region, and that there
should be a similar blue wing, then the absence of such a blue wing in the
observed spectrum might be explained by absorption. Partial absorption of the
blue wing due to an outflow is also seen in the Ly$\alpha$ line of hydrogen, for
example \citet{kriss2011}. However that absorption extends only up to
$-500$~km\,s$^{-1}$ in the UV spectrum. Here we would need absorption at several
thousands of km\,s$^{-1}$. To absorb the blue wing, we would need ionisation
stages containing \ion{Fe}{xviii} or higher, having 1s--2p transitions above
6.4~keV. However, application of our \textsl{xabs} models shows that this is
impossible: we would also get significant 1s--3p transitions at higher energies,
and the required column densities, of the order of $10^{28}$~m$^{-2}$ for
hydrogen, are inconsistent with the column densities that we derived from our
fits to the low-energy part of the spectrum. We conclude that if the red wing of
the narrow Fe-K emission line has an origin in an X-ray broad line region, then
the line must be asymmetric with significantly more flux in the red wing.

Another possibility is that the red wing is not a separate entity but part of a
broader structure, for instance a weak relativistic emission line. Indeed, we
get an excellent fit to the spectrum using the sum of the narrow component with
a relativistically broadened iron line. Such a line could stem from reflection
by relatively cool material. The line is weak, however, and therefore poorly
constrained, and the interpretation of the best-fit parameters must be done with
care. The inclination angle seems to be small, perhaps about 20 degrees, which
is consistent with a type 1 viewing direction, and with the fact that, for
instance, from the optical imaging of the \ion{O}{iii} lines
\citep{phillips1983} it appears as if we are looking fully through the
ionisation cone.

This relativistic line could be part of a more general reflection component. Our
best-fit reflection fraction $R$ of $0.64\pm 0.11$ then suggests that the
covering of this reflector is less than $2\pi$~sr.

From their analysis of the XMM-Newton pn data, \citet{ponti2013} also concluded
that the spectrum allows for a weak relativistic emission line. However, in
their case the relativistic line originates from \ion{Fe}{xxv} or \ion{Fe}{xxvi}
while we need colder iron. In the case of \citet{ponti2013} the data could be
modelled alternatively with ionised \ion{Fe}{xxv} and \ion{Fe}{xxvi} emission
lines. Our data are consistent with such a model.

\subsection{Ultra-fast outflows}

The HETGS spectrum shows no clear evidence of ultra-fast outflows in the form of
highly blueshifted absorption lines of \ion{Fe}{xxv} and \ion{Fe}{xxvi}. In
Sect.~\ref{subsect:fekregion} we showed that the presence of such lines is not
statistically significant at the expected blueshifts for an ultra-fast outflow.
However, given the statistical uncertainties we also cannot rule out such lines
at the strength of about 20~eV equivalent width as reported by
\citet{cappi2009}. For this equivalent width, we derive a hydrogen column
density of $1.2\times 10^{26}$ and $3.1\times 10^{26}$~m$^{-2}$ for
\ion{Fe}{xxv} and \ion{Fe}{xxvi}, respectively. These values are two orders of
magnitude higher than those measured for the ionisation components A to E, which
are of the order of a few times $10^{24}$~m$^{-2}$~(see Table~\ref{tab:fits}).
We can see in Fig.~\ref{fig:curpar0} that the measured hydrogen column density
of the absorbers in general increases with the ionisation parameter, a behaviour
often seen in other AGN \citep{behar2009}. However, for values of $\log\xi$ of
3.7 and 4.1 (for which the concentration of \ion{Fe}{xxv} and \ion{Fe}{xxvi}
peak, respectively) one would expect columns of the order of $\sim
10^{25}$~m$^{-2}$, much smaller than found for the ultra-fast outflows. This
supports the idea that the ultra-fast outflows, if present, may originate in a
different environment than the warm absorbers, possibly in winds launched close
to the accretion disc.

Similarly, troughs of \ion{Mg}{xii}, \ion{Si}{xiv} and \ion{S}{xv} outflowing at
$\sim 0.05c$ were marginally detected in the LETGS spectrum in the 5--9~\AA\
range \citep{ebrero2011}. This velocity was remarkably similar to that of the 
variable Fe absorption lines reported by \citet{ponti2009}, suggesting that both
absorbers might belong to the same ionised wind. While the lack of a significant
detection could be attributed to the low sensitivity of LETGS at those
wavelengths, one would expect to detect them significantly in the HETGS if they
are actually present. However, we did not find evidence of any of these
absorption features in our spectrum, obtaining tighter upper limits on their
column densities than those of \citet{ebrero2011}. These ions are typically
produced in a gas with ionisation parameter $\log\xi\sim 3$, very close to that
of component E2. The expected hydrogen column density is also similar to that
measured for component E2, which is significantly detected in our spectrum.
Therefore, if such an ionised wind was present in this observation it would have
been detected.

In summary, we do not find significant evidence of an ultra-fast outflow in this
Chandra HETGS observation of Mrk 509, in spite of previous detections by
\citet{cappi2009} and \citet{ponti2009}.  This result is in line with the
observation of no signatures of such outflows in the combined 600~ks XMM-Newton
pn spectrum, with the exception of a marginal detection of two absorption
features at 9~keV and 10.2~keV \citep{ponti2013} during observation 4 which
could be attributed to \ion{Fe}{xxvi} outflowing at $\sim -0.3c$. However, the
lack of detection of such outflows during this campaign may not rule out their
existence. Ultra-fast outflows are thought to originate very close to the
accretion disc, at distances typically of hundreds of Schwarzschild radii
\citep{tombesi2012}. Time variability is therefore an important structural
feature of these winds that may cause them to change face dramatically at
timescales of hours and days, thus preventing them from being detected in all
observations. More sensitive observations (e.g. ASTRO-H) are needed to get a
more conclusive view on ultra-fast outflows.

\section{Conclusions}\label{sect:conclusions}

In this paper we report a Chandra HETGS observation of Mrk 509 taken in
September 2012. The analysis of the HETGS spectrum confirms the structure of the
warm absorber measured with XMM-Newton RGS in 2009 by \citet{detmers2011}, who
found 5 distinct ionisation components in two kinematic regimes. 

We find no evidence of an outflow component at $-715$~km\,s$^{-1}$ as reported
by \citet{detmers2011} for the RGS spectrum. Most likely a few statistical
outliers at the blue wing of the \ion{Mg}{xi} and \ion{Fe}{xxi} lines caused the
apparent blueshift in the RGS spectrum, although we cannot exclude that is was a
transient feature.

We find no significant variability in the physical properties of the absorbers
between the 2009 and 2012 observations, except for the components C2 and D2 at
$-240$~km\,s$^{-1}$, where the column density of the lower ionisation component
C2 decreased in favour of the higher ionisation component D2.

The analysis of the line profiles of the \ion{O}{viii} Ly-series shows evidence
that the covering factor is less than unity. We obtain an average covering
factor of $0.81\pm 0.08$ for outflow velocities faster than $-100$~km\,s$^{-1}$,
similar to the values found in the UV for \ion{C}{iv} and \ion{N}{v} lines. This
can be interpreted as a patchy wind that leaks part of the underlying continuum,
although the significance of $f_{\rm c}<1$ is just above the $2\sigma$ level.

We also modelled the Fe-K emission. The parameters of the narrow component are
consistent with those determined by \citet{ponti2013}, and its steadiness in
flux between 2009 and 2012 points out to an origin distant from the central
engine. We see a strong red wing at the low energy side of the narrow line.
It is too strong for a Compton shoulder, and it is unlikely to be a symmetric
broad line that is completely absorbed on the blue side. Rather, it points
towards an intrinsically asymmetric broadened line, for instance a weak
relativistic line.

Our HETGS spectrum shows no evidence of ultra-fast outflows in the Fe-K region
nor in the 3~keV region. The lack of detection of these features, previously
reported in other Mrk~509 observations \citep{dadina2005,cappi2009,ponti2009},
might be due to intrinsic time variability, as they are formed very close to the
accretion disc.

\begin{acknowledgements}

This work is based on observations obtained with XMM-Newton, an ESA science
mission with instruments and contributions directly funded by ESA Member States
and the USA (NASA). It is also based on observations with INTEGRAL, an ESA
project with instrument and science data centre funded by ESA member states
(especially the PI countries: Denmark, France, Germany, Italy, Switzerland,
Spain), Czech Republic, and Poland and with the participation of Russia and the
USA. This work made use of data supplied by the UK Swift Science Data Centre at
the University if Leicester. SRON is supported financially by NWO, the
Netherlands Organization for Scientific Research. J.S. Kaastra thanks the PI of
Swift, Neil Gehrels, for approving the TOO observations.  M. Mehdipour
acknowledges the support of Studentship Enhancemement Program (STEP) awarded by
the UK Science \& Technology Facilities Council (STFC). N. Arav and G. Kriss
gratefully acknowledge support from NASA/XMM-Newton Guest Investigator grant
NNX09AR01G. Support for HST Program numbers 12022 and 12916 was provided by NASA
through grants from the Space Telescope Science Institute, which is operated by
the Association of Universities for Research in Astronomy, Inc., under NASA
contract NAS5-26555. E. Behar was supported by a grant from the ISF. S. Bianchi,
M. Cappi, and G. Ponti acknowledge financial support from contract ASI-INAF n.
I/088/06/0. P.-O. Petrucci acknowledges financial support from CNES and from the
Franco-Italian PICS of the CNRS. G. Ponti acknowledges support via EU Marie
Curie Intra-European Fellowships under contract no. FP7-PEOPLE-2009-IEF-331095
and FP-PEOPLE-2012-IEF-331095. 
\end{acknowledgements}

\bibliographystyle{aa}
\bibliography{paper}

\begin{thebibliography}{34}
\expandafter\ifx\csname natexlab\endcsname\relax\def\natexlab#1{#1}\fi

\bibitem[{{Arav} {et~al.}(2012){Arav}, {Edmonds}, {Borguet}, {Kriss},
  {Kaastra}, {Behar}, {Bianchi}, {Cappi}, {Costantini}, {Detmers}, {Ebrero},
  {Mehdipour}, {Paltani}, {Petrucci}, {Pinto}, {Ponti}, {Steenbrugge}, \& {de
  Vries}}]{arav2012}
{Arav}, N., {Edmonds}, D., {Borguet}, B., {et~al.} 2012, \aap, 544, A33

\bibitem[{{Arav} {et~al.}(2002){Arav}, {Korista}, \& {de Kool}}]{arav2002}
{Arav}, N., {Korista}, K.~T., \& {de Kool}, M. 2002, \apj, 566, 699

\bibitem[{{Behar}(2009)}]{behar2009}
{Behar}, E. 2009, \apj, 703, 1346

\bibitem[{{Cappi} {et~al.}(2009){Cappi}, {Tombesi}, {Bianchi}, {Dadina},
  {Giustini}, {Malaguti}, {Maraschi}, {Palumbo}, {Petrucci}, {Ponti},
  {Vignali}, \& {Yaqoob}}]{cappi2009}
{Cappi}, M., {Tombesi}, F., {Bianchi}, S., {et~al.} 2009, \aap, 504, 401

\bibitem[{{Chartas} {et~al.}(2003){Chartas}, {Brandt}, \&
  {Gallagher}}]{chartas2003}
{Chartas}, G., {Brandt}, W.~N., \& {Gallagher}, S.~C. 2003, \apj, 595, 85

\bibitem[{{Chartas} {et~al.}(2002){Chartas}, {Brandt}, {Gallagher}, \&
  {Garmire}}]{chartas2002}
{Chartas}, G., {Brandt}, W.~N., {Gallagher}, S.~C., \& {Garmire}, G.~P. 2002,
  \apj, 579, 169

\bibitem[{{Crenshaw} {et~al.}(1999){Crenshaw}, {Kraemer}, {Boggess}, {Maran},
  {Mushotzky}, \& {Wu}}]{crenshaw1999}
{Crenshaw}, D.~M., {Kraemer}, S.~B., {Boggess}, A., {et~al.} 1999, \apj, 516,
  750

\bibitem[{{Crenshaw} {et~al.}(2003){Crenshaw}, {Kraemer}, \&
  {George}}]{crenshaw2003}
{Crenshaw}, D.~M., {Kraemer}, S.~B., \& {George}, I.~M. 2003, \araa, 41, 117

\bibitem[{{Dadina} {et~al.}(2005){Dadina}, {Cappi}, {Malaguti}, {Ponti}, \& {de
  Rosa}}]{dadina2005}
{Dadina}, M., {Cappi}, M., {Malaguti}, G., {Ponti}, G., \& {de Rosa}, A. 2005,
  \aap, 442, 461

\bibitem[{{Detmers} {et~al.}(2011){Detmers}, {Kaastra}, {Steenbrugge},
  {Ebrero}, {Kriss}, {Arav}, {Behar}, {Costantini}, {Branduardi-Raymont},
  {Mehdipour}, {Bianchi}, {Cappi}, {Petrucci}, {Ponti}, {Pinto}, {Ratti}, \&
  {Holczer}}]{detmers2011}
{Detmers}, R.~G., {Kaastra}, J.~S., {Steenbrugge}, K.~C., {et~al.} 2011, Paper
  III, \aap, 534, A38

\bibitem[{{Ebrero} {et~al.}(2011){Ebrero}, {Kriss}, {Kaastra}, {Detmers},
  {Steenbrugge}, {Costantini}, {Arav}, {Bianchi}, {Cappi},
  {Branduardi-Raymont}, {Mehdipour}, {Petrucci}, {Pinto}, \&
  {Ponti}}]{ebrero2011}
{Ebrero}, J., {Kriss}, G.~A., {Kaastra}, J.~S., {et~al.} 2011, Paper V, \aap,
  534, A40

\bibitem[{{Elvis}(2000)}]{elvis2000}
{Elvis}, M. 2000, \apj, 545, 63

\bibitem[{{Kaastra} {et~al.}(2011{\natexlab{a}}){Kaastra}, {de Vries},
  {Steenbrugge}, {Detmers}, {Ebrero}, {Behar}, {Bianchi}, {Costantini},
  {Kriss}, {Mehdipour}, {Paltani}, {Petrucci}, {Pinto}, \&
  {Ponti}}]{kaastra2011b}
{Kaastra}, J.~S., {de Vries}, C.~P., {Steenbrugge}, K.~C., {et~al.}
  2011{\natexlab{a}}, Paper II, \aap, 534, A37

\bibitem[{{Kaastra} {et~al.}(2012){Kaastra}, {Detmers}, {Mehdipour}, {Arav},
  {Behar}, {Bianchi}, {Branduardi-Raymont}, {Cappi}, {Costantini}, {Ebrero},
  {Kriss}, {Paltani}, {Petrucci}, {Pinto}, {Ponti}, {Steenbrugge}, \& {de
  Vries}}]{kaastra2012}
{Kaastra}, J.~S., {Detmers}, R.~G., {Mehdipour}, M., {et~al.} 2012, Paper VIII,
  \aap, 539, A117

\bibitem[{{Kaastra} {et~al.}(2000){Kaastra}, {Mewe}, {Liedahl}, {Komossa}, \&
  {Brinkman}}]{kaastra2000}
{Kaastra}, J.~S., {Mewe}, R., {Liedahl}, D.~A., {Komossa}, S., \& {Brinkman},
  A.~C. 2000, \aap, 354, L83

\bibitem[{{Kaastra} {et~al.}(1996){Kaastra}, {Mewe}, \&
  {Nieuwenhuijzen}}]{kaastra1996}
{Kaastra}, J.~S., {Mewe}, R., \& {Nieuwenhuijzen}, H. 1996, in UV and X-ray
  Spectroscopy of Astrophysical and Laboratory Plasmas, ed. {K.~Yamashita \&
  T.~Watanabe}, 411

\bibitem[{{Kaastra} {et~al.}(2011{\natexlab{b}}){Kaastra}, {Petrucci}, {Cappi},
  {Arav}, {Behar}, {Bianchi}, {Bloom}, {Blustin}, {Branduardi-Raymont},
  {Costantini}, {Dadina}, {Detmers}, {Ebrero}, {Jonker}, {Klein}, {Kriss},
  {Lubi{\'n}ski}, {Malzac}, {Mehdipour}, {Paltani}, {Pinto}, {Ponti}, {Ratti},
  {Smith}, {Steenbrugge}, \& {de Vries}}]{kaastra2011a}
{Kaastra}, J.~S., {Petrucci}, P.-O., {Cappi}, M., {et~al.} 2011{\natexlab{b}},
  Paper I, \aap, 534, A36

\bibitem[{{Kazanas} {et~al.}(2012){Kazanas}, {Fukumura}, {Behar},
  {Contopoulos}, \& {Shrader}}]{kazanas2012}
{Kazanas}, D., {Fukumura}, K., {Behar}, E., {Contopoulos}, I., \& {Shrader}, C.
  2012, The Astronomical Review, 7, 030000

\bibitem[{{Kriss} {et~al.}(2011){Kriss}, {Arav}, {Kaastra}, {Ebrero}, {Pinto},
  {Borguet}, {Edmonds}, {Costantini}, {Steenbrugge}, {Detmers}, {Behar},
  {Bianchi}, {Blustin}, {Branduardi-Raymont}, {Cappi}, {Mehdipour}, {Petrucci},
  \& {Ponti}}]{kriss2011}
{Kriss}, G.~A., {Arav}, N., {Kaastra}, J.~S., {et~al.} 2011, Paper VI, \aap,
  534, A41

\bibitem[{{Krolik} \& {Kriss}(2001)}]{krolik2001}
{Krolik}, J.~H. \& {Kriss}, G.~A. 2001, \apj, 561, 684

\bibitem[{{Lodders} \& {Palme}(2009)}]{lodders2009}
{Lodders}, K. \& {Palme}, H. 2009, Meteoritics and Planetary Science
  Supplement, 72, 5154

\bibitem[{{Magdziarz} \& {Zdziarski}(1995)}]{magdziarz1995}
{Magdziarz}, P. \& {Zdziarski}, A.~A. 1995, \mnras, 273, 837

\bibitem[{{Matt}(2002)}]{matt2002}
{Matt}, G. 2002, \mnras, 337, 147

\bibitem[{{Phillips} {et~al.}(1983){Phillips}, {Baldwin}, {Atwood}, \&
  {Carswell}}]{phillips1983}
{Phillips}, M.~M., {Baldwin}, J.~A., {Atwood}, B., \& {Carswell}, R.~F. 1983,
  \apj, 274, 558

\bibitem[{{Pinto} {et~al.}(2012){Pinto}, {Kriss}, {Kaastra}, {Costantini},
  {Ebrero}, {Steenbrugge}, {Mehdipour}, \& {Ponti}}]{pinto2012}
{Pinto}, C., {Kriss}, G.~A., {Kaastra}, J.~S., {et~al.} 2012, Paper IX, \aap,
  541, A147

\bibitem[{{Ponti} {et~al.}(2013){Ponti}, {Cappi}, {Costantini}, {Bianchi},
  {Kaastra}, {De Marco}, {Fender}, {Petrucci}, {Kriss}, {Steenbrugge}, {Arav},
  {Behar}, {Branduardi-Raymont}, {Dadina}, {Ebrero}, {Lubi{\'n}ski},
  {Mehdipour}, {Paltani}, {Pinto}, \& {Tombesi}}]{ponti2013}
{Ponti}, G., {Cappi}, M., {Costantini}, E., {et~al.} 2013, Paper XI, \aap, 549,
  A72

\bibitem[{{Ponti} {et~al.}(2009){Ponti}, {Cappi}, {Vignali}, {Miniutti},
  {Tombesi}, {Dadina}, {Fabian}, {Grandi}, {Kaastra}, {Petrucci}, {Bianchi},
  {Matt}, {Maraschi}, \& {Malaguti}}]{ponti2009}
{Ponti}, G., {Cappi}, M., {Vignali}, C., {et~al.} 2009, \mnras, 394, 1487

\bibitem[{{Steenbrugge} {et~al.}(2011){Steenbrugge}, {Kaastra}, {Detmers},
  {Ebrero}, {Ponti}, {Costantini}, {Kriss}, {Mehdipour}, {Pinto},
  {Branduardi-Raymont}, {Behar}, {Arav}, {Cappi}, {Bianchi}, {Petrucci},
  {Ratti}, \& {Holczer}}]{steenbrugge2011}
{Steenbrugge}, K.~C., {Kaastra}, J.~S., {Detmers}, R.~G., {et~al.} 2011, paper
  VII, \aap, 534, A42

\bibitem[{{Tombesi} {et~al.}(2012){Tombesi}, {Cappi}, {Reeves}, \&
  {Braito}}]{tombesi2012}
{Tombesi}, F., {Cappi}, M., {Reeves}, J.~N., \& {Braito}, V. 2012, \mnras, 422,
  L1

\bibitem[{{Tombesi} {et~al.}(2013){Tombesi}, {Cappi}, {Reeves}, {Nemmen},
  {Braito}, {Gaspari}, \& {Reynolds}}]{tombesi2013}
{Tombesi}, F., {Cappi}, M., {Reeves}, J.~N., {et~al.} 2013, \mnras, 430, 1102

\bibitem[{{Tombesi} {et~al.}(2010){Tombesi}, {Cappi}, {Reeves}, {Palumbo},
  {Yaqoob}, {Braito}, \& {Dadina}}]{tombesi2010}
{Tombesi}, F., {Cappi}, M., {Reeves}, J.~N., {et~al.} 2010, \aap, 521, A57

\bibitem[{{Yaqoob} \& {Padmanabhan}(2004)}]{yaqoob2004}
{Yaqoob}, T. \& {Padmanabhan}, U. 2004, \apj, 604, 63

\bibitem[{{Zycki} \& {Czerny}(1994)}]{zycki1994}
{Zycki}, P.~T. \& {Czerny}, B. 1994, \mnras, 266, 653

\bibitem[{{Zycki} {et~al.}(1999){Zycki}, {Done}, \& {Smith}}]{zycki1999}
{Zycki}, P.~T., {Done}, C., \& {Smith}, D.~A. 1999, \mnras, 305, 231

\end{thebibliography}

\end{document}